\documentclass[a4paper,byrevtex,nofootinbib,showkeys,dvips]{revtex4}
\usepackage{dcolumn}
\usepackage{amsmath}
\usepackage{amssymb}
\usepackage{bm}

\begin{document}

\title{Do bras and kets have dimensions?}

\author{Claude \surname{Semay}}
\email[E-mail: ]{claude.semay@umons.ac.be}
\thanks{ORCiD: 0000-0001-6841-9850}

\author{Cintia \surname{Willemyns}}
\email[E-mail: ]{cintia.willemyns@umons.ac.be}
\thanks{ORCiD: 0000-0001-8114-0061}

\affiliation{Service de Physique Nucl\'{e}aire et Subnucl\'{e}aire,
Universit\'{e} de Mons,
UMONS Research Institute for Complex Systems,
Place du Parc 20, 7000 Mons, Belgium}

\date{\today}
\begin{abstract}
The bra and ket notation introduced by Dirac and the dimensional analysis are two powerful tools for the physicist. Curiously, almost nothing is said about connections between these two topics in the literature. We show here that bras and kets have dimensions. This could help students to grasp a better comprehension of this abstract notation. 
\end{abstract}


\maketitle

The bra and ket notation introduced by Dirac is widely used in many textbooks (see \cite{griff18} for instance). Although not devoid of problems and shortcomings \cite{gier00}, this formalism is very powerful to perform changes of basis or to handle spinor states, for instance. The usefulness of the dimensional analysis must no longer be demonstrated \cite{soni01,misi10}, whether it is to simply check formulas or to build new physical laws in difficult problems. It can be particularly suited in quantum mechanics which is not known for its very intuitive nature \cite{karp20}. Nevertheless, very few information can be found about the dimensions of bras and kets in the literature. This is not a comfortable situation. Since bras and kets are relevant physical quantities, they are expected to be characterized by units or dimensions \cite{maha18}. The following text is a detailed analysis of a question found on the Internet: ``Do bras and kets have dimensions?" \cite{int}. It is shown that the question is relevant and that the answer is positive. This could help students to grasp a better comprehension of this abstract notation.

In this paper, we will determine the dimensions of bras and kets for a one-body system in one dimension. But the results can be easily extended to $N$-body systems in $D$ dimensions. In the following, we will only use the dimensions of mass, length and time: $M$, $L$ and $T$. This is convenient for our purpose but there is some freedom in the choice of the basic dimensions \cite{soni01,levy98}. As usual, the dimension of a quantity $Q$ is noted $[Q]$, with the convention $[Q]=1$ if $Q$ is dimensionless. If $x$ is the position and $p$ the conjugate momentum, then
\begin{equation}
\label{xphbar}
[x]=L \quad \textrm{and} \quad [p]=M\,L\, T^{-1}, 
\end{equation}
with $[\hbar]=[x]\,[p]=M\,L^2\, T^{-1}$. If $| \alpha \rangle$ and $| \beta \rangle$ are kets of the Hilbert space considered, then
\begin{equation}
\label{dimfond}
[\langle \alpha  | \beta \rangle] = [\langle \alpha  |]\, [| \beta \rangle] \quad \textrm{and} \quad [\langle \alpha  | \beta \rangle] = [\langle \beta  | \alpha \rangle].
\end{equation}
The first relation stems from the fact the bracket $\langle \alpha  | \beta \rangle$ is the result of the action of the bra $\langle \alpha  |$ on the ket $| \beta \rangle$. The second one is the result of the equality $\langle \alpha  | \beta \rangle = \langle \beta  | \alpha \rangle^*$ since the complex conjugation does not change the dimensions. But there is a priori no reason to assume that $[| \alpha \rangle]=[\langle \alpha |]$. We will come back to this point below. If the ket $| \psi \rangle$ represents a physical state, let us note $[| \psi \rangle]=K$ and $[\langle \psi |]=B$, where $K$ and $B$ are dimensional formulas. A physical state is always normalisable (but it could be built by a superposition of non-normalisable states like the wave packet of a free particle) \cite{griff18}. With relations~(\ref{dimfond}), the normalisation condition $\langle \psi  | \psi \rangle = 1$ implies that $B=K^{-1}$. 

The continuous position basis $\{ | x \rangle\}$ is such that $\hat x\, | x \rangle = x\, | x \rangle$, where $\hat x$ is the position operator. The wave function in the position space $\Psi(x,t)$ is defined by \cite{griff18}
\begin{equation}
\label{x1}
\langle x  | \Psi \rangle = \langle \Psi  | x \rangle^* = \Psi(x,t), 
\end{equation}
with $[\Psi(x,t)] = L^{-1/2}$ because of the normalisation condition
\begin{equation}
\label{normx}
\int_{-\infty}^{+\infty} |\Psi(x,t)|^2\,dx =1.
\end{equation}
Let us note $[\langle x  |]=X_B$ and $[| x \rangle]=X_K$. Equation~(\ref{x1}) implies
\begin{equation}
\label{dimx1}
X_B\, K = X_K\, K^{-1} =  L^{-1/2}.
\end{equation}
The closure relation is given by \cite{griff18}
\begin{equation} 
\label{x2}
| \psi \rangle = \int_{-\infty}^{+\infty} \Psi(y,t)\,| y \rangle\,dy \quad \textrm{with} \quad 
\langle x  | y \rangle = \delta(x-y),
\end{equation}
where $y$ is a position. The last equality with the Dirac delta function is called ``Dirac orthonormality" in \cite{griff18}. Equations~(\ref{x2}) imply respectively\,\footnote{It is our teaching experience that students are generally not aware that wave functions and Dirac delta function carry dimensions. So, it is useful to insist on this point.}
\begin{equation}
\label{dimx2}
K=X_K\, L^{1/2} \quad \textrm{and} \quad X_B\, X_K = L^{-1} .
\end{equation}
This does not bring any new information, since (\ref{dimx2}) can be deduced from (\ref{dimx1}). So, at this stage, relations exist between $X_B$, $X_K$ and $K$, but their dimensional formulas are not fixed.  

Let us look at the continuous momentum basis $\{ | p \rangle\}$. It is such that $\hat p\, | p \rangle = p\, | p \rangle$, where $\hat p$ is the momentum operator with $[\hat x,\hat p]=i\,\hbar$. The wave function in the momentum space $\Phi(p,t)$ is defined by \cite{griff18}
\begin{equation}
\label{p1}
\langle p  | \Psi \rangle = \langle \Psi  | p \rangle^* = \Phi(p,t). 
\end{equation}
$\Phi(p,t)$ is normalised because $\Psi(x,t)$ is normalised \cite{griff18}. If we note $[\langle p  |]=P_B$ and $[| p \rangle]=P_K$, the normalisation condition implies
\begin{equation}
\label{dimp1}
P_B\, K = P_K\, K^{-1} =  (M\,L\, T^{-1})^{-1/2}.
\end{equation}
The corresponding closure relation, given by \cite{griff18}
\begin{equation} 
\label{p2}
| \psi \rangle = \int_{-\infty}^{+\infty} \Phi(q,t)\,| q \rangle\,dq \quad \textrm{with} \quad 
\langle p  | q \rangle = \delta(p-q),
\end{equation}
where $q$ is a momentum, does not bring supplementary information about dimensions, as in the case of the position basis. 

Using the Dirac notation, it is very easy to find the link between the position-space wave function $\Psi(x,t)$ and the momentum-space wave function $\Phi(p,t)$: \cite{griff18} 
\begin{equation}
\label{PhiPsi}
\Phi(p,t) = \int_{-\infty}^{+\infty} \langle p  | y \rangle \, \Psi(y,t)\, dy, 
\end{equation}
where
\begin{equation}
\label{xp}
\langle p  | x \rangle = \langle x  | p \rangle^* = \frac{1}{\sqrt{2\,\pi\,\hbar}}e^{-i\,p\,x/\hbar}. 
\end{equation}
Simple manipulations show that (\ref{PhiPsi}) and (\ref{xp}) are compatible with (\ref{dimx1}) and (\ref{dimp1}), but with nothing new. The definition of bases relying on other conjugate variables will lead to the same results. 

But a physical state $| \Psi \rangle$ can also be expanded in a discrete orthonormal basis $\{| n \rangle\}$, \cite{griff18}
\begin{equation} 
\label{n1}
| \Psi \rangle = \sum_{n=1}^{\infty} C_n(t)\,| n \rangle,
\end{equation}
with
\begin{equation} 
\label{n2}
C_n(t) =  \langle n  | \Psi \rangle \quad \textrm{and} \quad \langle n | m \rangle = \delta_{n m}.
\end{equation}
The normalisation condition gives
\begin{equation} 
\label{normn}
\langle \Psi| \Psi \rangle = \sum_{m,n=1}^{\infty} C_m(t)^*\,C_n(t)\,\langle m | n \rangle=1.
\end{equation}
This implies $[C_m(t)^*]\,[C_n(t)]\,[\langle m | n \rangle]=[C_n(t)]^2\,[\langle m | n \rangle]=1$, that is to say $[C_n(t)]=1$, thanks to the normalisation condition for the discrete states. Noting $[|n\rangle]=N_K$ and $[\langle n|]=N_B$, (\ref{n1}) and its ``bra" counterpart give immediately
\begin{equation}
\label{NKNB}
K=N_K \quad \textrm{and} \quad B=N_B=K^{-1}.
\end{equation}
In fact, this is not very surprising since the basis states $| n \rangle$ are generally the eigenstates of a given operator $\hat A$, $\hat A\, | n \rangle = a_n | n \rangle$, with a definite physical content.

So, the dimensions of the various bras and kets presented above cannot be univocally determined. The attribution of dimensions to a quantity is to some extent a matter of convention, provided the global coherence of the whole system is guaranteed \cite{soni01,levy98}. It is then not shocking to fix a convention for bras and kets. Two conditions appear quite naturally. 

First, in order to simplify all the relations given above, it can be decided that $[|\Psi\rangle]=K=1$. As a consequence, a ket and its corresponding bra for a physical state are dimensionless. This is in complete agreement with the Dirac notation, which is by construction independent of any representation. But this is not the case for the continuous basis states which refer explicitly to a given representation.

Second, we can safely claim that there can be no difference between the physics with kets or with bras. For instance, the counterpart of (\ref{x2}) is 
\begin{equation} 
\label{bra}
\langle \psi|  = \int_{-\infty}^{+\infty} \Psi(y,t)^*\,\langle y| \,dy.
\end{equation}
It seems then also quite natural to enforce that the dimensions for a ket and its corresponding bra are equal. 

Fortunately, the examination of formulas above shows that these two conventions are equivalent:
\begin{equation}
\label{equiv}
[\langle \alpha  |]=[ | \alpha \rangle] \Leftrightarrow K=1.
\end{equation}
So,
\begin{equation}
\label{dimxp}
K = B = N_K = N_B = 1, \quad X_B = X_K = L^{-1/2} \quad \textrm{and} \quad P_B = P_K = (M\,L\, T^{-1})^{-1/2}.
\end{equation}
We conclude that the notion of dimensions is relevant for bras and kets. Our results are resumed in Table~\ref{tab:1}. From these properties, $[\langle \alpha  | \beta \rangle] = [\langle \alpha  |]\, [| \beta \rangle]= [| \alpha  \rangle]\, [\langle \beta |]= [\langle \beta  | \alpha \rangle]$. This is compatible with the fact that the complex conjugation does not change the dimensions. 
\begin{table}[hbt]
\begin{center}
\caption{\label{tab:1} Dimensions for bras and kets for a one-body system in one dimension. The two equivalent conventions are marked in bold.}
\begin{tabular}{|l|l|}
\hline
general property & $[\langle \alpha  | \beta \rangle] = [\langle \alpha  |]\, [| \beta \rangle]$ \\
\textbf{bra/ket} & $[\langle \alpha  |]=[ | \alpha \rangle]$ \\
\textbf{physical state} & $[ |\Psi \rangle] =1$ \\
position basis & $[ |x \rangle] =L^{-1/2}$ \\
momentum basis & $[ |p \rangle] =(M\,L\, T^{-1})^{-1/2}$ \\
discrete basis & $[ |n \rangle] =1$ \\
\hline
\end{tabular}
\end{center}
\end{table}

All the results obtained above are easy to generalise for different situations. For a $N$-body system in a space with $D$ dimensions, $\Psi(\bm r_1,\ldots,\bm r_N) = \langle \bm r_1,\ldots,\bm r_N | \Psi \rangle$ and $[\Psi(\bm r_1,\ldots,\bm r_N)]= L^{-ND/2}$. If a basis ket is associated with a dimensioned continuous variable, its dimension is determined by the corresponding Dirac orthonormality. For a discrete basis, the corresponding kets are dimensionless.

Finally, it could seem strange that the dimensions of a ket change with its status. But we can attempt a (bold) parallel with the expansion of a three-dimensional vector $\bm a = \sum_{i=1}^3 a_i\, \bm e_i$ in various bases. In a Cartesian basis, $[\bm e_x]=[\bm e_y]=[\bm e_z]=1$, while in a spherical natural basis, $[\bm e_\rho]=1$ and $[\bm e_\theta]=[\bm e_\phi]=L$. This is obvious by examining the corresponding metric in which the dimensions of the basis elements $\bm e_i$ appear through the values of $\bm e_i^2$.  

\begin{acknowledgments}
One of the author (CS) would like to thank Thomas Brihaye for drawing his attention to the existence of reference \cite{gier00}, which was the starting motivation for this paper. This work was supported by the Fonds de la Recherche Scientifique - FNRS under Grant Number 4.4510.08. 
\end{acknowledgments}

\end{document}